\RequirePackage{fix-cm}
\documentclass[smallextended]{svjour3}       
\usepackage[utf8]{inputenc}
\usepackage[english]{babel}
 
\usepackage{hyperref}
\usepackage[utf8]{inputenc}
\usepackage[english]{babel}
\usepackage{amssymb,amsmath, mathtools}
\usepackage{natbib}
\setcitestyle{aysep={}}
\usepackage{color}
\usepackage{pdfsync}
\usepackage{url}
\usepackage{subfigure}
\usepackage{bbold}
\usepackage{dsfont}
\usepackage{epsfig}
\usepackage{epstopdf}
\usepackage{newlfont}
\usepackage{multirow}
\usepackage{longtable} 
\usepackage{ifthen}
\usepackage{alltt}
\usepackage{enumerate}
 \usepackage[text={6.25in,8.5in},centering]{geometry}

\newcommand{\veps}{\varepsilon}

\newcommand{\beq}{\begin{eqnarray*}}
\newcommand{\feq}{\end{eqnarray*}}
\newcommand{\beqn}{\begin{eqnarray}}
\newcommand{\feqn}{\end{eqnarray}}
\newcommand{\bec}{\begin{claim}}
\newcommand{\fec}{\end{claim}}
\newcommand{\becn}{\begin{claim*}}
\newcommand{\fecn}{\end{claim*}}

\newtheorem{assume}[theorem]{Assumption}

\begin{document}

\title{Phase Transitions in a Logistic Metapopulation Model with Nonlocal Interactions
}


\author{Ozgur Aydogmus 
}


\institute{Ozgur Aydogmus\at
              Social Sciences University of Ankara, Oran/Ankara/Turkey \\
              Tel.: +90-312-596-4416\\
              \email{ozgur.aydogmus@asbu.edu.tr}
}

\date{Received: date / Accepted: date}
\maketitle

\begin{abstract}
The presence of one or more species at some spatial locations but not others is a central matter in ecology. This phenomenon is related to ecological pattern formation. { Nonlocal interactions can be considered as one of the mechanisms causing such a phenomenon. We propose a single-species, continuous time metapopulation model taking nonlocal interactions into account.  Discrete probability kernels are used to model these interactions  in a patchy environment. A linear stability analysis of the model shows that solutions to this equation exhibit pattern formation if the dispersal rate of the species is sufficiently small and the discrete interaction kernel satisfies certain conditions. We numerically observe that traveling and stationary wave type patterns arise near critical dispersal rate.} We use weakly nonlinear analysis to better understand the behavior of formed patterns. We show that observed patterns arise
through both supercritical and subcritical bifurcations from spatially homogeneous steady state. Moreover, we observe that as the dispersal rate decreases, amplitude of the patterns increases. For discontinuous transitions to instability, we also show that there exists a threshold for the amplitude of the initial condition, above which pattern formation is observed.

\end{abstract}


\section{Introduction}
\label{intro}

As noted by \citet{durlev}, one of the fundamental issues of modeling spatially structured ecological systems can be considered as the choice of level of detail. For spatial ecological processes, the spatiotemporal dynamics of the system can be described using different spatial scales. \cite{durlev} compared four approaches to model spatially distributed systems which can be summarized as follows: mean field dynamics, reaction-diffusion equations (RDE), patch dynamics and interacting particle systems. We are interested in analyzing an intraspecific metapopulation model through the paper. Yet, we also summarize recent developments concerning RDEs (or more generally space-continuous models) in this section.

Continuum models such as RDEs, integro-differential or integro-difference equations are used to describe ecological processes widely (see e.g. \citet{book,holmesPDE,lewis,kot92}). Our particular interest here is in pattern formation mechanisms of ecological processes. After Turing's seminal work on morphogenesis \citep{turing}, pattern formation in reaction-diffusion systems has been an intensive area of research in ecology (see, for example, \cite{segel72,okubo}). Additionally, conditions for generating spatially distributed patterns for systems of integro-difference equations were studied by \cite{kot} 

An important mechanism generating patterns is nonlocal interactions. \citet{main,main1} criticized the assumption that growth of a species in a spatial habitat depends only on the local population density. Instead he assumed that consumption of resources at a spatial location does not solely depend on the local population density but also on the population's weighted average at that point. For some variants of nonlocal Fisher equation, it was shown that the spatially homogenous solution becomes unstable when nonlocal competition is taken into account \citep{pattern,ben,kuperman1,kuperman2,doubly}. The effect of nonlocal interactions for integro-difference equations  was also studied by \citet{ben2}. These models are basically  spatially-extended versions of logistic model. There are also a number of other studies that incorporated nonlocal competition to the systems of RDEs \citep{symmetric,asymmetric,volpert2,banerjee}.

Here we would like to indicate that the main mathematical tool for studying pattern formation in continuous space equations is Fourier transform and/or Fourier series. When we have this well-established analytical technique in hand, we can go beyond the linear stability analysis and ask about phase transitions between different behaviors. This  technique is called ``weakly nonlinear analysis'', informing us about transitions to instabilities near a bifurcation point \citep{murray}. Such an analysis allows the classification of possible bifurcations. {At a supercritical bifurcation there is a smooth transition from the stable zero amplitude solution to the stable bifurcating branch as the bifurcation parameter approaches its critical value and the zero solution loses stability. Whereas, the zero solution loses stability at a subcritical bifurcation as the bifurcation parameter approaches its critical value, yet there is no small amplitude
stable solution and so the amplitude of the solutions must jump from zero to some
other stable solution at large amplitude \citep{patternbook}. Thus supercritical and subcritical bifurcations are associated with continuous and discontinuous changes in amplitudes, respectively.} Weakly nonlinear analysis was used for an animal aggregation model \citep{lewis2} to understand how transitions between disordered and ordered behaviors depend on the population density or if these transitions exhibit hysteresis. As a result, it was reported that solutions to the equation exhibit discontinuous transitions ({\it i.e.} subcritical bifurcation) from ordered to disordered motion. Weakly nonlinear analysis of another continuum model for animal aggregation was also performed and subcritical bifurcations were reported {near a bifurcation point} \citep{lewis1} . Recently a similar analysis was employed for nonlocal Fisher equation and it is  was shown that both subcritical and supercritical bifurcations are possible near the critical diffusion rate \citep{ben}.

In the following we take a landscape ecology perspective and subdivide the environment into distinct patches each of which is defined as a homogenous geographic region. Dispersal between these patches are defined as movement from a breeding site (or a birth site) to another breeding site \citep{dispersal}. Following \citet{book}, these movements are modeled using discrete probability kernels. For a review of multi-patch models coupled via dispersal between patches see \cite{hanski,hastings,gilpin, cosner12, book}. 

{Here our aim is to study a continuous time, logistic metapopulation model which incorporates nonlocal effects. As in the case of dispersal, we choose to model nonlocal interactions using another discrete probability kernel determining the interaction range and strength. The motivation for the study of nonlocal interactions is related to the competition for common resources.  Resources such as food, energy or water are basic requirements for the growth of a population. Competition for these resources in neighboring spatial locations through foraging as in ant colonies \citep{ant} or consumption of diffusible resources such as water by plants \citep{veg1,veg2,metanl} can be considered as examples of nonlocal interactions.

Other types of nonlocal interactions have also been studied in the literature. For example, a species of land (or amphibious) animals that live on the shores of a lake was studied by \citet{canadian}. In their model the patches correspond to segments of the shoreline forming a ring and the nonlocal interaction is due to the accumulation of
population's waste production in the the lake. Lastly, one can consider signalling between cells as another example of nonlocal interactions. Juxtacrine cell signalling is an important cellular communication mechanism between cells \citep{owen} playing an important role in spatial pattern generation and wound healing process \citep{monk}. Such a mechanism is modeled by \citet{owen}  via coupled differential equations with the assumption that interactions between neighboring cells are nonlinear.}

We would like to note that a spatial extension of logistic model, discrete in space and time, was studied by \cite{doebelli}, and it  was reported that quasi-local interactions destabilize the space homogenous equilibrium of the model. The assumption of quasi-local interactions implies that every patch interacts only with their neighboring patches, and thus the linearization of such a system leads us to a toeplitz matrix. As a result, the eigenvalues of the system can be studied numerically. For other works exploring the effect of quasi-local interactions in metapopulations, see \cite{utz,kisdi}. Note also that a similar analysis can be repeated for continuous time metapopulation models taking quasi-local interactions into account.

In our model, we assume that a species lives in a patchy environment containing $N$ uniformly distributed sites with wrap around boundaries. Such an assumption is justified by considering, for example, that the habitats lie along the shore of a lake or pond. In this setting, we also assume that the movement and competition for resources between two patches depend on the distance between them, yet we do not require symmetric or `local' interactions in any sense. The resulting model is a nonlinear $N$-dimensional system of ordinary differential equations. To be able to deal with the problem of stability of space homogenous solution to this system, we use discrete Fourier series (DFS). From the results of this analysis, we conclude that the nonlocal interactions may destabilize the space homogenous equilibrium of the system. We also illustrate numerical simulations of traveling and stationary wave type patterns near the critical diffusion rate.

As mentioned above, having a tool like DFS enables us to ask questions related to the phase transitions. Thus we use classical Stuart Landau (S-L) theory \citep{stuart} to study the effect of nonlocal interactions near the stability boundary. We derive cubic S-L equations governing the behavior of the solutions for a large period of time. Using the cubic equation, we can classify possible bifurcations when the system parameters are known.  As a result we find  that both supercritical and subcritical bifurcations may arise near stability boundaries. In the case of continuous transitions to instability, we approximate the amplitudes of the patterns using the stable equilibrium of cubic S-L equations. In the case of discontinuous transitions to instability, the cubic equation implies the existence of a threshold. If the population's initial configuration has an amplitude below this threshold, spatial inhomogeneity will vanish. Whereas, we observe the formation of patterns persisting for a long period of time if the initial amplitude is above this threshold. However, the cubic equation does not allow us to approximate the amplitudes of patterns when the model exhibits subcritical bifurcations. In order to obtain more information about these patterns it is also important to have these approximations. We derive quintic S-L equations and consequently can obtain such approximations. Our numerical investigation shows that the nonlinear prediction stays close to the computationally obtained amplitudes near stability boundaries. For both types of bifurcations we observe that decreasing the bifurcation parameter (dispersal rate) near its critical value results in more denser groups of individuals with higher amplitudes.

The paper is organized as follows: In Section \ref{modelnotation}, we start by introducing the notation of the paper and detailing our model. We perform a linear stability analysis of the model near its spatially homogenous solution and find conditions to observe pattern formation in Section \ref{LA} where numerical simulation of stationary and traveling wave type patterns are also illustrated. In Section \ref{NLA}, we obtain cubic and quintic S-L equations for our metapopulation model. We conclude the paper in Section \ref{conc}.

\section{The model and the notation of the paper}\label{modelnotation}

In the following subsection, we introduce and fix the notation of the paper. In addition, we give some properties of DFS. The latter subsection is devoted to model formulation. 

\subsection{The discrete Fourier series}

In developing our model, we consider our habitat as a portion of the earth on which species are able to colonize and live. Considering a habitat that lie along a lake or pond and subdividing it into distinct patches enables us to consider a one dimensional periodic habitat $S.$  {We label each site in this habitat by an integer and denote the set of these integers by ${S}:=\{0,1,\cdots, N-1\}.$ We denote by $p_n$ the population density at site n and introduce the vector notation $\mathbf p:=(p_n)_{n\in S}=(p_{0},p_{1},\cdots,p_{N-1})$.} Following \citet{dft,mandal}, {discrete Fourier series (DFS)} of this vector is denoted by $\mathcal F\mathbf p=\mathbf P=(P_{k})_{k\in S}$ and given by \beq P_k=\sum_{n\in S}p_n e^{{-2j\pi kn}/{N}}\feq for any $k\in S.$ Note that $j$ is used to denote the imaginary unit number.

Define the componentwise multiplication (or Hadamart product) of two vectors $\mathbf x$ and $\mathbf y$ as 

\beq \mathbf x\circ \mathbf y:= (x_{n}y_{n})_{n\in S}.\feq 
Thus in accordance with this definition of multiplication, any power of vector $\mathbf x$ can be defined by 

\beq \mathbf{x}^\alpha:= (x_{n}^\alpha)_{n\in S}.\feq
Additionally, we consider the vector of complex exponentials {$\mathbf{w} = \bigl(e^{2\pi j n/N}\bigr)_{n\in S}.$} For integers $m_1$ and $m_2,$ we have the following orthogonality property of complex exponential vectors:
\beq
\mathbf{w}^{m_1}\cdot\mathbf{w}^{m_2}=\begin{cases}
N ~~\text{ if } m_1 \equiv m_2 ~~~\text{modulo}~ N \\
0 ~~~\text{ if } m_1 \not\equiv m_2 ~~~\text{modulo}~ N
\end{cases}
\feq where $\cdot$ is the inner product of two vectors in $\mathbb{C}^{N}.$ This property follows from the discussion by \citet[p. 104]{dft}. Moreover, we would like to note that the set $\{\mathbf w^m|m=0,1,\cdots,N-1\}$ is a basis for the vector space $\mathbb{C}^{N}.$

Lastly, denote the discrete convolution operator of two vectors $\mathbf x$ and $\mathbf y$ by $\mathbf x*\mathbf y$ and define it as follows: 

\beq (\mathbf x*\mathbf y)_i=\sum_{n\in S} x_iy_{<i-n>_N}\feq where ${<i-n>_N}=i-n ~\text{modulo}~ N. $ It can be seen that the DFS of the convolution of two vectors $\mathbf x$ and $\mathbf y$ is given by the Hadamart product of DFS of these vectors {\it i.e.} we have:

 \beq \mathcal F{(\mathbf x*\mathbf y)}=\mathbf X\circ\mathbf Y. \feq 
For a review of these results, see \cite{dft,mandal}.

\subsection{The model}

Consider our one dimensional periodic habitat $S.$ If an individual with site of origin $n$ adopts another site $i$ for residence, this movement is called the dispersal from site $n$ to site $i$. One can chose to model such movements  between discrete patches of this habitat by discrete dispersal or island chain models \citep{levin,hastings2,book}.  Such models can be set in either discrete or continuous time \citep{book}. Following \cite{levin} and \cite{hastings2}, we consider a continuous time model. Suppose that in each patch the population grows or declines according to a differential equation $dp_n/dt = f_n(\mathbf{p})$ where $\mathbf{p}=(p_n)_{n\in S}.$ Next we describe how the population density $p_i$ on each patch $i\in S$ varies in time.

Here we consider passive dispersal of individuals between patches which is quantified in terms of two major parameters: (1) {\it dispersal rate} is the expected proportion of individuals  to leave a patch and (2) {\it dispersal range or distance} is usually described by a discrete dispersal kernel which gives the probability distribution of the distance traveled by an individual to find a new breeding site \citep{dispersal}. 
Suppose that individuals disperse from patch $n$ at a rate of $\delta_n\geq 0$ and arrive at patch ${i}$ at a rate of ${\tilde d_{in} \geq 0}.$ Hence our model is given as follows:

 \beqn\label{geneq} \frac{dp_n}{dt}={\sum_{i\in S}\tilde d_{ni}p_i} -\delta_np_n +f_n(\mathbf{p})\feqn 
with $\delta_n={\sum_{i\in S}\tilde d_{ni}}$ which implies that there is no mortality in transit. Suppose also that the dispersal rates depend on the distance between two sites {$i$} and $n$ {\it i.e.} we have: 

\beq {\tilde d_{ni}=\bar d_{<n-i>_N}}.\feq
 Hence, the summation in equation \eqref{geneq} defines a circular convolution as follows: 

\beq (\mathbf d*\mathbf p)_n:= {\sum_{i\in S}\bar d_{<n-i>_N}p_i}.\feq 
This also implies that the dispersal rates satisfies $\delta:=\delta_n$ for any $ n\in S.$

\cite{hastings2,levin} assumed that the growth rate of the population at a site depends only on the population density at the same patch. However, \citet{main} and \citet{doebelli} criticized this assumption and  claimed that the population growth depends not only on the resources in the local patch but also on the resources in neighboring spatial locations. {One can consider, for instance, situations in which reproduction and hence population dynamics  take place in a habitat patch whose resources are also used by individuals that live
and reproduce in neighboring patches through foraging. Therefore the depletion of resources in a given habitat patch occurs not only due to the individuals living and reproducing in this patch, but also due to individuals living in a certain neighborhood of that patch. Vegetation dynamics is another interesting example in which nonlocal competition appears to
be occurring \citep{veg1,veg2,metanl}. Since water is a diffusible resource it follows that the effect of a plant consuming water locally is  that it consumes water from the surrounding regions. Thus the greater the rate of water uptake  by a given biomass at a patch, the lower the quantity of water available at neighboring patches implying that nearby patches with low soil moisture may be hostile for a new agent. In this example, the vegetation dynamics is governed by local
resource uptake, yet nonlocal competition results from the diffusion of the resource. Hence, we consider nonlocal coupling between habitat patches taking into account not only dispersal between neighboring patches but also nonlocal interactions. }

Taking into account research conducted by \citet{main} and \citet{doebelli}, we assume that the resources available to an individual is given by the spatial average of the resources in a certain neighborhood of the breeding site of the individual. Considering logistic growth and following the model setup presented by \cite{pattern} lead us to the following growth rate of the population at site $n$:  

\beqn\label{fn} f_n(\mathbf{p})=p_n\bigg(a_n-(\mathbf c*\mathbf p)_n\bigg)\feqn
where $a_n$ is the carrying capacity of site $n$ and $\mathbf c$ is the nonlocal competition kernel. \cite{japan} analyzed the continuous space analog of this equation with slowly varying carrying capacity in space variable and concluded that the model exhibits very complicated dynamics. Since we want to analyze the behavior of patterns in terms of their amplitudes, we have a simplifying assumption that is given by $a_n=1$ for all $n\in S.$ This is to say that each patch is identical; hence, $\mathbf p=\mathbf1$ is the space homogenous solution of the equation.

Thus, we get the following system of differential equations 

\beqn \label{int}\frac{dp_n}{dt}=\delta\bigl((\mathbf d*\mathbf p)_n -p_n\bigr) +p_n\bigl(1-(\mathbf c*\mathbf p)_n\bigr)\feqn
where both $\mathbf d$ and $\mathbf c$ are discrete probability distributions on $\mathbb{Z}$ (or simply on $S$). {The functions $f_n$ defined in \eqref{fn} are non-dimensionalized for time; in other words, equation \eqref{int} has been scaled in time so that $\delta$ is the relative time scale between dispersal and reproduction.} We remark that the above given model mimics the nonlocal Fisher equation \citep{main,gourley} in a patchy environment.

We summarize the above given assumptions regarding our model as follows:

\begin{itemize}
\item[(A1)] {Regarding the dispersal kernel $\mathbf d$, we have assumed that it is a probability distribution on $S$ describing the proportion of individuals dispersing from one site to another by accounting the distance between them. Biologically, this implies that individuals are not lost or do not reproduce while dispersing. In addition, unless the kernel is degenerate ({\it i.e.} $d_0=1$) the descendants of an individual from a site $i$ can be found in any site in $S$ after enough generations. The dispersal between two spatial locations is assumed to depend on the distance between them due to the limited mobility of individuals.}

\item[(A2)] {Regarding the competition or interaction kernel $\mathbf c,$ we have assumed that it satisfies the same assumptions with the dispersal kernel. This kernel was used to calculate average frequency of the species, competing for the common resources, in a neighborhood of any spatial location. Since competition is restricted to a neighborhood, it is also realistic to assume that this kernel takes the spatial distance between sites into account.  }

\item[(A3)] {Regarding the reproduction term $f_n(\mathbf p)$, we have assumed that it is a modified logistic growth function. We would like to note that this growth term can be obtained by specifying birth and death rates of the species (see \citet[section 2]{pattern}). Biologically, this growth term implies that  the more offspring are produced in a patch  as the population gets
larger in the site. Whereas the reproduction and survival decrease in a site as the average density of individuals increases in the neighborhood of the site. The latter is due to the fact that the carrying capacity affects only the average population in the neighborhood of any patch in our model. This also implies that the population size may exceed the carrying capacity in some spatial patches provided that the average population size is small enough. }

\end{itemize}

\section{Linear Analysis and Pattern Formation}\label{LA}

We perform a linear stability analysis of the model to examine the effect of the dispersal rate $\delta$ and dispersal and interaction kernels $\mathbf{d}$ and $\mathbf{c}$ on destabilization of the space homogenous equilibrium $\mathbf1$ of \eqref{int}. In addition, we give two numerical examples to illustrate traveling and stationary wave-type patterns arising from \eqref{int}.

\subsection{Linear Stability Analysis}\label{LSA}
If we assume that there is no spatial effect {\it i.e.} {$d_0= c_0=1$} , \eqref{int} becomes the simple logistic equation with stable solution $p_n=1$ for all $n\in S.$ 

To investigate necessary conditions for pattern formation, we linearize \eqref{int} by using the first order expansion $\mathbf{p}=1+\veps \check{\mathbf p}e^{\lambda t},$ where $\check{\mathbf p}$ is a spatial perturbation term independent of time $t.$ {Hence this formula can be considered as a separation of variables formula for \eqref{int}. Substituting this ansatz to the equation gives the following first order relation:}
\begin{equation}
\lambda \check{\mathbf p}=\delta(\mathbf d*\check{\mathbf p}-\check{\mathbf p}) - \mathbf c*\check{\mathbf p}.
\end{equation}

By taking the DFS of both sides, one gets

\begin{equation}
\label{eigen}
\lambda (\delta,k)=\delta(D_k-1) - C_k
\end{equation} where $\mathbf{D}=(D_k)_{k\in S}$ and $\mathbf{C}=(C_k)_{k\in S}$ are DFS of probability kernels $\mathbf d$ and $\mathbf c,$ respectively.
Here we would like to note that the Fourier transforms (or characteristic functions) of these kernels can take complex values since the kernel functions are not necessarily symmetric. Therefore, we introduce notations $\mathcal{R}[z]$ and $\mathcal{I}[z]$ to denote real and imaginary parts of a complex number $z,$ respectively. 

Note that stability condition for the equilibrium $\mathbf1$ is given as $\mathcal{R}[\lambda(\delta,k)]<0$ for all $k\in S.$ Since kernel $\mathbf{d}$ is a discrete probability distribution, {$\mathcal{R}\big[{\mathbf D}-\mathbf1\big]\leq 0$} for the dispersal kernel. This implies that we need to require $\mathcal{R}[{ C_k}]< 0$ for some $k\in S$ to destabilize steady state $\mathbf p=\mathbf 1.$ {The effects of nonlocal dispersal and competition are now more apparent. Increasing the dispersal rate between patches flattens the spatial heterogeneity (or it may have no effect on the stability of the equilibrium, see Assumption \ref{ass}). This is to say that the spatial inhomogeneity arises as a result of nonlocal averaging. }

{Define the set $\tilde S=\{k\in S: \mathcal{R}[{ C_k}]< 0\}.$ If the DFS of the dispersal kernel $D_k$ is equal to 1 for some $k\in \tilde S$ then the state $\mathbf 1$ is unstable for any positive dispersal rate $\delta.$ To proceed and define the critical dispersal rate we need the following assumption. 

\begin{assume}\label{ass}
$D_k\neq 1$ for any $k\in \tilde S.$
\end{assume} 
Hereafter we always require that Assumption \ref{ass} holds. {This assumption is not a requirement to observe spatially inhomogeneous solutions but it rules out some kernels for which dispersal loses its flattening effect. Hence it guarantees the existence of a positive dispersal rate at which spatial homogeneity vanishes. This rate will be used as a bifurcation point for the further analysis of the model.} Note that this assumption is not redundant, yet the discussion is rather technical and given in Appendix \ref{discussion}. 

One can easily conclude that steady state $\mathbf{p}=\mathbf 1$ is unstable for $\delta=0$ as long as $\mathcal{R}[C_k]<0$ for some $k\in S.$ By continuity, it is also unstable for small enough $\delta>0.$ By Assumption \ref{ass}, $\mathcal{R}[D_k- 1]< 0$ for any $k\in\tilde S.$ Hence, for large enough $\delta>0$ the state will be stable. The critical dispersal rate $\delta_0$ is calculated by solving the following maximization problem: }

\beqn\label{maxim} \max_{n\in S} \mathcal R[ \lambda (\delta, n)] = 0.\feqn
We denote the argument of the maximization problem by {$k_c$} which is called the critical wavenumber. We remark that the critical wave number $k_c$ is in the set $\tilde S.$  

\subsection{Computed stationary and traveling waves}\label{comp}

Here we focus on instabilities around stability boundaries. Our numerical simulations show that only {stationary and traveling waves can be observed near stability boundaries (for an illustration of these patterns see Figure \ref{stationary} Panels (a) and (b), respectively ).}

It has been shown that the solutions to {nonlocal Fisher equation} exhibit pattern formation if the real part of the characteristic function of nonlocal contact term takes negative values \citep{pattern}. Hence, symmetric uniform kernels have been used widely to model nonlocal interactions (see for example \citet{kuperman1,kuperman2,pattern}). Following recent works by \citet{symmetric,asymmetric,ben}, we assume that each convolution term depends on two parameters. $r_d$ and $a_d$ characterize the range of nonlocality and extent of asymmetry for dispersal kernel $\mathbf d,$ respectively. Similarly, $r_c$ and $a_c$ characterize the range of nonlocality and extent of asymmetry for interaction kernel $\mathbf c,$ respectively. Taking $a_i=0$ for $i=d$ or $c$ implies that the corresponding kernel is symmetric. {As an example, one can consider the family of step functions for positive integers $a$ and $r$ as follows: 

\begin{equation}
\label{kernel}
v_n^{(r,a)} =
\begin{cases}
\frac{1}{2r+1} & \text{if } |n-a| \leq r \\
0 & \text{otherwise .} 
\end{cases}
\end{equation} for $n \in \mathbb Z.$} Note that equation \eqref{kernel}  is an infinite sequence which is denoted by $\mathbf{v}_\infty^{(r,a)}:=\big(v_n^{(r,a)}\big)_{n\in\mathbb{Z}}.$

\begin{figure}[h!]

\centering

\includegraphics[width=16cm, height=6.5cm]{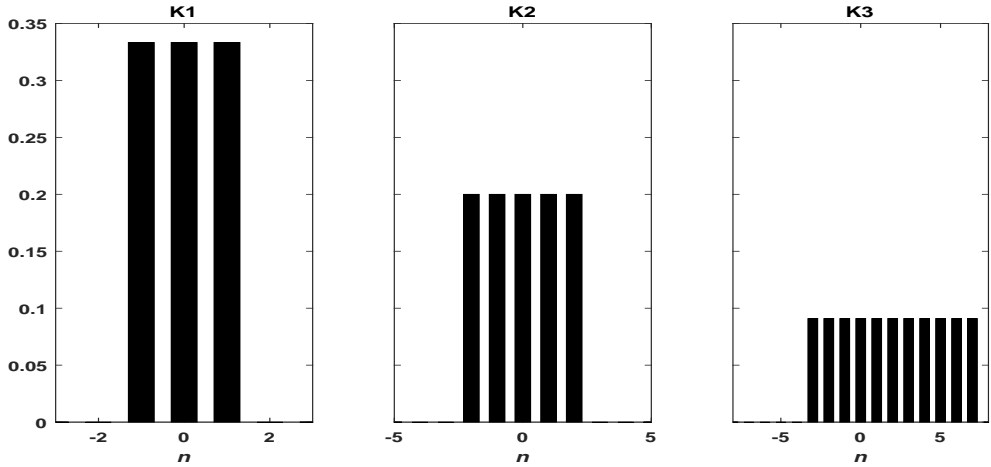}
\caption{Examples of discrete uniform kernels. Panel (K1) illustrates uniform kernel $\mathbf v_\infty^{(1,0)}$ which will be used as both dispersal and interaction kernels to simulate standing wave type patterns. Similarly, Panel (K2) shows a discrete top-hat kernel $\mathbf v_\infty^{(2,0)}$ modeling next nearest neighbor coupling \citep{metanl}. Lastly interaction kernel $\mathbf v_\infty^{(5,2)}$ shown in Panel (K3) will be employed to simulate asymmetric nonlocal interactions. }
\label{kernels}
\end{figure}

For the sake of simplicity, we assume assume  that $2r+1\leq N.$ {Hence, \eqref{kernel} can be considered as an $N$ dimensional vector which is denoted by $\mathbf u^{(r,a)}$ with $u_n^{(r,a)}:=v_{<n>_N}^{(r,a)}.$} The DFS of the uniform kernel $\mathbf u^{(r,a)}$ is calculated in Appendix \eqref{dfsu}. {We remark that discrete uniform kernel \eqref{kernel} is a discretization of the top-hat kernel used to model nonlocal dispersal for continuous space ecological processes \citep{tophat}. }

We consider two examples for {$N={17}$} as follows:
\begin{itemize}
\item[\textbf{E1:}] Stationary waves, in Figure \ref{stationary} (a), were obtained for the parameters $\mathbf d=\mathbf c=\mathbf u^{(1,0)}$ (see Figure \ref{kernels} Panel (K1)), initial condition $1+0.001\cos(k_c x)$ and $\delta=0.2435$ while the critical value of the nonlocal dispersal rate is given by $\delta_0=0.2436.$
\item[\textbf{E2:}] Figure \ref{stationary} (b) shows periodic traveling waves obtained for the kernel functions $\mathbf d=\mathbf u^{(2,0)}$ and $\mathbf c=\mathbf u^{(5,2)}$ (see Figure \ref{kernels} Panels (K2) and (K3), respectively), initial condition $1+0.001\cos(k_c x)$ and the dispersal rate is taken as the critical nonlocal dispersal rate {\it i.e.} $\delta=\delta_0=0.1103.$ 
\end{itemize}

{Our biological motivation for considering these two examples are as follows. In example \textbf{E1}, we assumed that competition for resources occurs not only between individuals in a given patch, but also to some extent between individuals reproducing in neighboring patches to which individuals may leave their home territory in search of food. In the second example $\textbf{E2},$ on the other hand we assumed that the interaction range is larger than the dispersal range. Juxtacrine cell signaling can be considered as an example for which nearest neighborhood signaling mechanism account for the observed increase in cell proliferation many cell diameters away from the wound edge \cite{monk}. For some plant populations, this assumption might also be relevant. In the case of vegetation dynamics studied by \citet{veg1}, for example, the range of inhibitory effect of nonlocal water consumption is assumed to be larger than the dispersal distance. In the same study it was also assumed that the interaction kernel is isotropic {\it i.e.} slope imposing to rainwater is a well-defined running down direction. We can relax this assumption and consider an environment with anisotropy. This relaxation leads us to consider asymmetric nonlocal competition which may be modeled by using a discrete asymmetric kernel (see, Figure \ref{kernels} Panel (K3)). }

	Figure 1 (a) shows the stationary wave-type patterns obtained using parameters given in \textbf{E1}.  Figure 1 (b), on the other hand, shows periodic traveling waves that were obtained for the kernel functions and parameters described in \textbf{E2}. {It is clear from the Figure \ref{stationary} that at some patches the population level exceeds the carrying capacity $\mathbf 1$ while at some others it remains below the carrying capacity as expected (see Section \ref{modelnotation}, Assumption (A3)). This is to say that individuals form clusters or groups. We would like to note that standing wave-type patterns arises when both interaction and dispersal kernels are symmetric and these groups does not move in space as time passes. If the model takes asymmetric dispersal or interaction into account then traveling wave-type patterns arises and these clusters moves in space. }

\begin{figure}[h!]

\centering
\subfigure[]{
\includegraphics[width=6cm, height=6.5cm]{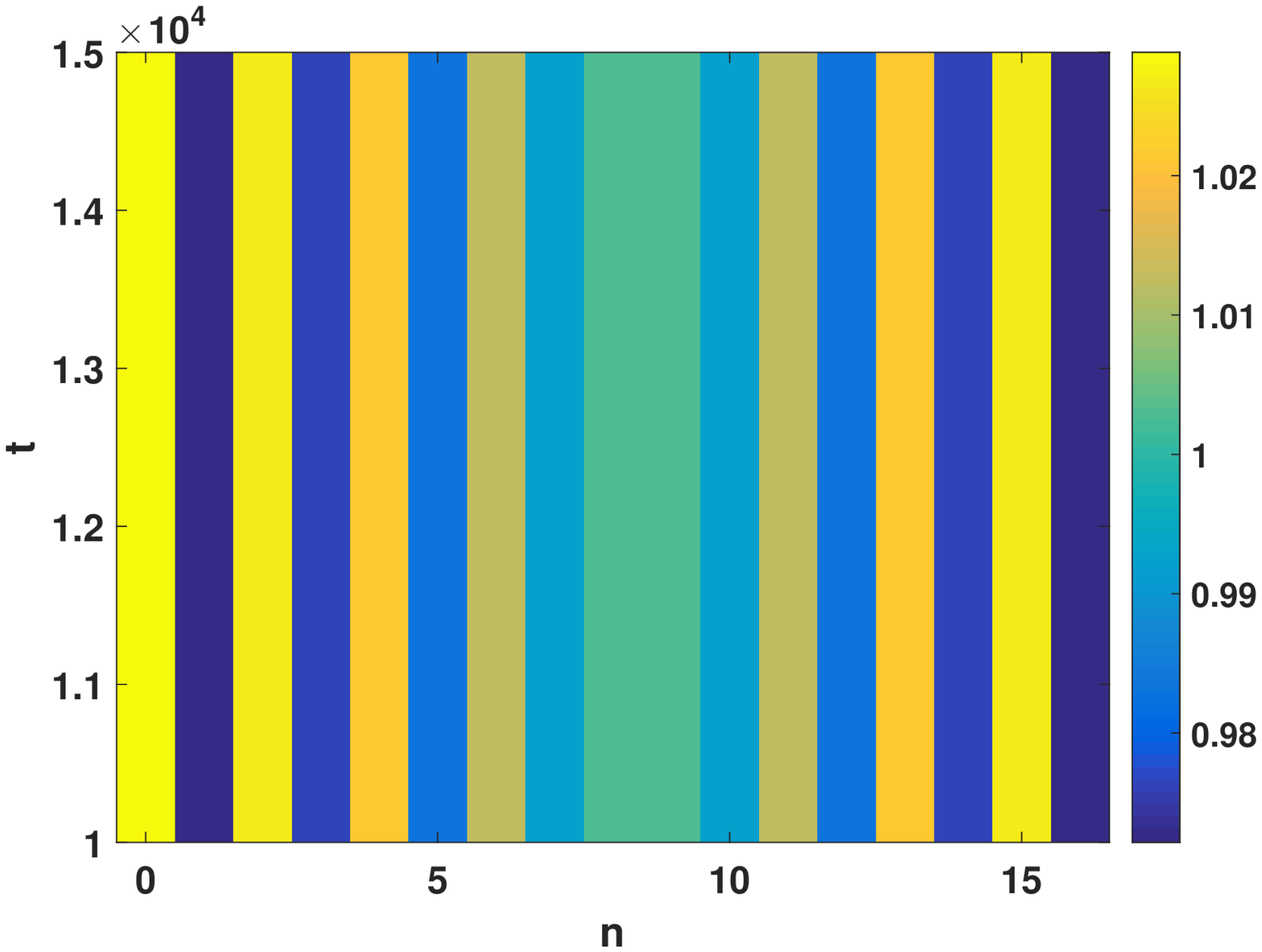}
}
\subfigure[]{
\includegraphics[width=6cm, height=6.5cm]{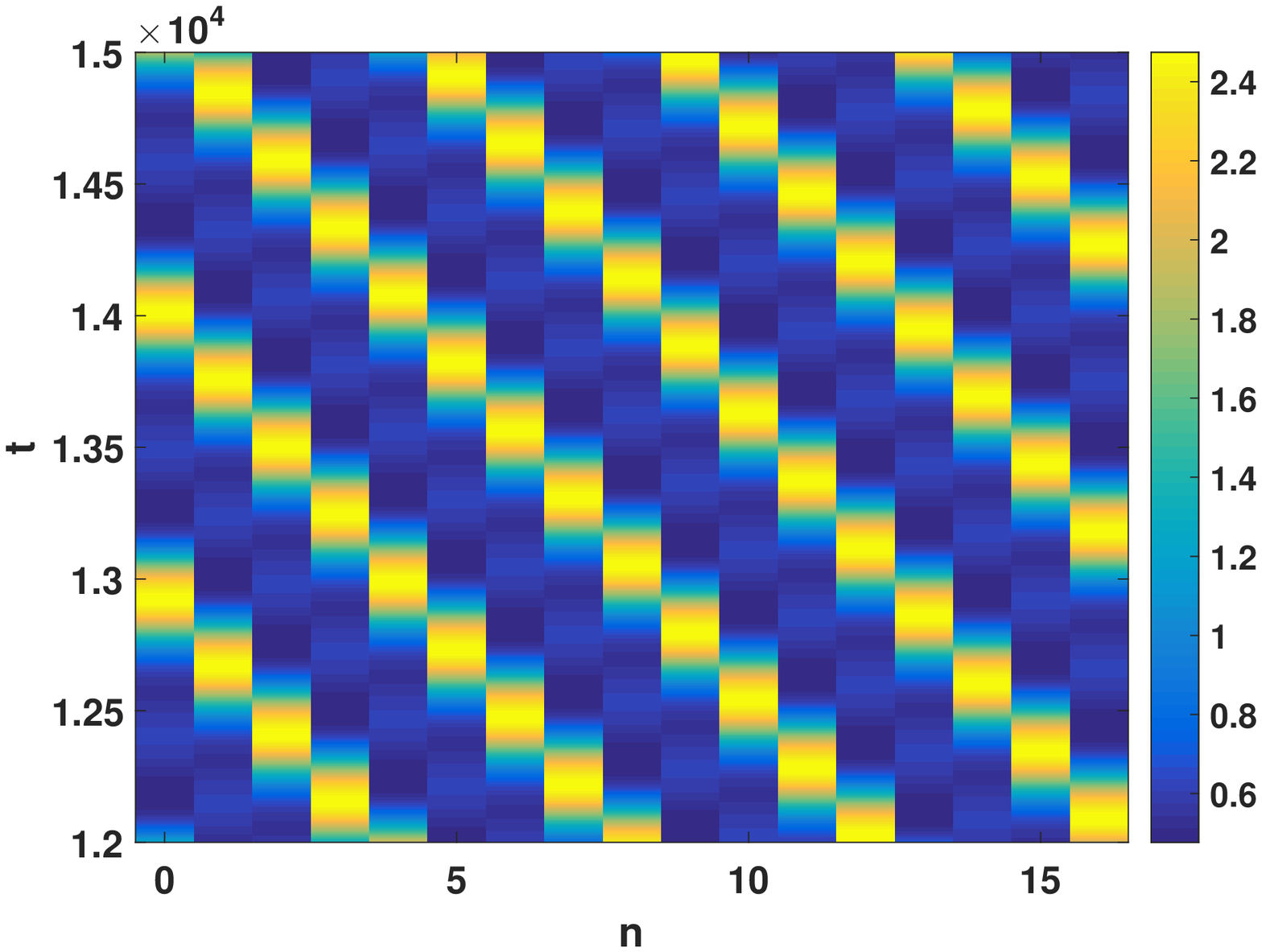}
} 
\caption{Stationary and traveling waves. Panel (a) illustrates amplitude level sets of stationary waves for the parameter set described in \textbf{E1}. Similarly, panel (b) shows traveling waves for the parameter set given in \textbf{E2}. }
\label{stationary}
\end{figure}

{Here, we would like to call attention to another significant difference between Figure \ref{stationary} (a) and (b). Recall the discussion on supercritical and subcritical bifurcations in Section \ref{intro}. From the color bars of these figures, it is possible to see that the amplitudes of stationary and traveling wave solutions are qualitatively different from each other even though each dispersal rate is very close to the corresponding critical dispersal rate. Define the half amplitude of equation \eqref{int} as follows:

\beqn\label{amp} \text{Amp} :=\frac{\max_{n\in S}\{\mathbf p(t^*)\}-\min_{n\in S}\{\mathbf p(t^*)\}}{2}\feqn for a sufficiently large $t^*.$  For the stationary waves, we numerically obtained the amplitude of the solution $\text{Amp}^{sw} =0.058.$ Similarly, we got $0.9684\leq\text{Amp}^{tw}\leq0.9975$ for the traveling waves amplitude of which oscillates in time (see Figure 4, Panel (b)). Our numerical results, therefore, indicate that continuous and discontinuous transitions to instability for the parameters given in \textbf{E1} and \textbf{E2}, respectively. To further investigate the qualitative difference between transitions to instability, we use weakly nonlinear analysis in Section \ref{NLA}.}

\section{Weakly Nonlinear Analysis}\label{NLA}

The linear stability analysis is only valid for small time and infinitesimal perturbations. For a large time, nonlinear terms affect the growth of unstable modes. In this section, we consider model \eqref{int} and obtain the {Stuart-Landau (S-L)} equation that gives information about the qualitative behavior of the solutions  {\citep{stuart}.} Our particular aim in this section is to explain the reason why amplitudes of the solutions $\text{Amp}^{sw}$ and $\text{Amp}^{tw}$ are significantly different from each other. On the other hand, searching for an answer to this question enables us to answer some other questions frequently asked concerning space-continuous models (see, for example, Section \ref{intro} or \cite{lewis2}). 

We would like to highlight a technical restriction regarding the validity of the results obtained from weakly nonlinear analysis. Since our model is a finite dimensional coupled ODE system, the set $\{\mathbf w^{mk_c}:m\in S\}$ may or may not constitute a basis for $\mathbb C^N.$ This problem is related to the algebraic structure of {the periodic habitat} $S=\mathbb Z_N$ with multiplication modulo $N.$ Hence we need some technical restrictionsıns to obtain S-L equations. These restrictions are given in the Appendices \ref{cubic} and \ref{ap} as Assumptions \ref{a1} and \ref{a2}.

\subsection{Cubic S-L equation and characterization of transitions to instabilities}\label{analytical}

We are interested in the stability of the homogenous solution $\mathbf p=\mathbf 1$ near the critical nonlocal dispersal rate $\delta_0$ with periodic boundary conditions. {We aim to find a solution to weakly nonlinear approximation to \eqref{int}} by multiple-scales perturbation expansion. Before proceeding to this expansion, we consider a perturbation of the bifurcation parameter $\delta$ as 
\beqn\delta=\delta_0+\veps^2 \mu \feqn 
for $0<\veps\ll 1$ and $\mu=\pm 1.$ Here $\mu$ determines the direction of the deviation from the critical nonlocal dispersal rate $\delta_0.$ Note that $\mathbf p=\mathbf1$ is unstable if $\mu=-1$ by the linear stability analysis presented in Section \ref{LSA} 

Let $\delta_0$ be the critical dispersal rate for which we have $\lambda(\delta_0,k_c)= 0.$ The solution to the linearization of \eqref{int} around $\mathbf 1$ and near critical dispersal rate is given as follows:
 \beqn\label{sln}
\mathbf p\propto \mathbf w^{k_c}e^{\lambda(\delta,k_c)t}+c.c.
\feqn 
where $c.c.$ stands for complex conjugate and $\mathbf w$ is as defined in Section \ref{modelnotation}.

Denote the imaginary part of the eigenvalue associated with the critical wavenumber by $w_0:=\mathcal I[\lambda(\delta_0,k_c)].$  {Expanding the eigenvalue $\lambda(\delta,k_c)$ in power series leads to 

\beqn
\lambda(\delta,k_c)=iw_0+\nu\veps^2 (D_{k_c}-1)+O(\veps^4)\nonumber
\feqn 
where we used equality \eqref{eigen} to obtain $\partial \lambda/\partial \delta.$} By substituting this expansion to the above given solution \eqref{sln}, one obtains 

\beqn\mathbf w^{k_c}e^{\lambda(\delta,k_c)t}&=&\mathbf w^{k_c}e^{iw_0t+\nu\veps^2 (D_{k_c}-1)t} \nonumber\\
&=&A(\veps^2 t) \mathbf w^{k_c}e^{iw_0t}.\nonumber\feqn 
Hence amplitude $A$ is a function of the slow time $\tau:=\veps^2t.$ We consider the fast and slow time scales $t$ and $\tau$ together in our analysis {\it i.e.} $t\to t+\tau.$

{Consider the family of linear waves in discrete space as follows:}

 \beqn\label{solution} W_m=\mathbf w^{<mk_c>_N }e^{imw_0t}\feqn 
for $m\in S$ and assume the following formal expansion:

 \beqn\label{psoln} \mathbf p=1+\veps\mathbf p^{(1)}+\veps^2 \mathbf p^{(2)}+\veps^3 \mathbf p^{(3)}\feqn where \beq \mathbf p^{(1)}&=&A(\tau)W_1+ c.c.\\ \mathbf p^{(2)}&=&A_{20}(\tau)W_0 +(A_{22}W_2+c.c.)\\ \mathbf p^{(3)}&=&A_{31}(\tau)W_1 +A_{33}W_3+c.c.\feq 

\begin{remark}
One can take the perturbation expansion as $ \mathbf p=1+\veps^\eta\mathbf p^{(1)}+\veps^{2\eta}\mathbf p^{(2)}+\cdots.$
\begin{itemize}
\item The first expansion term comes from the solution of the linearized equation \eqref{sln}. At $O(\veps^{2\eta}),$ the terms $W_{\pm1}$ only couples with $W_0$ and $W_{\pm2}.$ Higher order correction terms can be determined using the same logic. 
\item {To balance the unbounded growth predicted by the linear stability analysis, the magnitudes of the nonlinear term $\mathbf f(\mathbf p)=(f_n(\mathbf p))_{n\in S}$ and $\mathbf p_\tau$ need to be comparable.} This implies that $\eta=1.$ 
\end{itemize}
\end{remark}

Plugging the solution \eqref{psoln} in \eqref{int} leads to the following cubic S-L equation: \begin{equation}
\label{sl}
A_\tau=\mu \Phi A+\Psi A|A|^2
\end{equation} 
where coefficients $\Phi$ and $\Psi$ are explicitly given by formulas \eqref{phi} and \eqref{psi} along with the details of the derivation of \eqref{sl} in Appendix \ref{cubic}. Note also that equation \eqref{sl} is valid only if Assumption \ref{a1} is satisfied.

Since solutions to amplitude equations \eqref{sl} are complex, one can redefine this amplitude function as $A(\tau)=L(\tau)e^{i\theta(\tau)}$ with real terms $L(\tau)=|A(\tau)|$ and $\theta(\tau).$ Hence, an equivalent form of equation \eqref{sl} is given as follows:

\begin{flalign}
\label{li}
\frac{dL}{d\tau}&=\mu\mathcal{R}[\Phi]L+\mathcal{R}[\Psi]L^3,\\
\frac{d\theta}{d\tau}&=\mu\mathcal{I}[\Phi]+\mathcal{I}[\Psi]L^2. \nonumber
\end{flalign}

Equilibria of the equation \eqref{li} are $L_0=0$ and $L_1=\sqrt{-\mu\mathcal{R}[\Phi]/\mathcal{R}[\Psi]}$ {provided that $\mathcal{R}[\Psi]\neq0.$ If this is not true then the amplitude equation \eqref{li} becomes linear. In this case, higher order correction terms are needed to approximate the amplitude levels of pattens.} One can easily see that the solution $0$ of \eqref{li} is stable whenever $\mu\mathcal{R}[\Phi]<0.$ Thus, it is clear from \eqref{phi} that equilibrium $\mathbf p=\mathbf1$ (or $L_0$) is unstable for $\mu=-1$ and it is stable for $\mu=1.$ This result is in accordance with the result of the linear analysis given in Section \ref{LSA} 

It is clear that the equilibrium $L_1$ needs to be real i.e. $\mu\mathcal{R}[\Psi]\geq0.$ Thus the sign of $\mathcal{R}[\Psi]$ determines the direction of deviation from bifurcation parameter. One can categorize the possible bifurcations arising from \eqref{li} as follows:
\begin{enumerate}
\item If $\mathcal{R}[\Psi]>0,$ then $\mu=-1$ and \eqref{li} describes a supercritical bifurcation with stable branch $\veps L_1$ to the left of $\delta_0.$ 
\item If $\mathcal{R}[\Psi]<0,$ then $\mu=1$ and \eqref{li} describes a subcritical bifurcation with unstable branch $\veps L_1$ to the right of $\delta_0.$ 
\end{enumerate}

Here, the first case implies that a supercritical pitchfork bifurcation arises from the constant density state $\mathbf p=\mathbf1$ to a spatially periodic pattern with a steady state amplitude determined by $L_1.$  In the second case, transitions to instabilities are discontinuous and $L_1$ determines the maximum amplitude of the initial conditions above which $\mathbf p=\mathbf 1$ is unstable. {In the case of supercritical transitions, it is clear that spatially periodic patterns arises only if the dispersal rate of the population is less than the critical dispersal rate. In addition, amplitude levels as a measure of the spatial fluctuations or cluster sizes increases as mobility of individuals decreases. In the case of subcritical bifurcations, on the other hand, we only obtained an unstable branch for the values of dispersal rates that are larger than the critical dispersal rate. We will investigate this case in more detail computationally and analytically in the remainder of this section.}

\subsubsection*{Computational validation of analytical results}

In the following lines, we consider \eqref{int} with kernels described in \textbf{E1-2}. Amplitude levels presented in Figure \ref{stationary} implies the existence of both subcritical and supercritical bifurcations. One can calculate the values of $\Psi$ for kernels presented in \textbf{E1-2} as $\Psi^{sw}=-0.6476$ and $\Psi^{tw}=0.0288$ by using the formula \eqref{psi}. Hence, by the results of Section \ref{analytical}, transition to instability for stationary waves is continuous while it is discontinuous for traveling waves.

In order to compare the amplitude levels obtained from simulations to our analytical findings, we need to write the quantity $\text{Amp}$ defined in \eqref{amp} in terms of the amplitude function $A.$ Substitute the perturbation extension \eqref{ansatz} $\mathbf p=1+\veps \mathbf p^{(1)}$ to \eqref{amp}. Following \citet{lewis2}, one can easily obtain 

\beqn\label{amplitude} \text{Amp}= f(t^*)\veps \mathcal R[A] \feqn 
where $f(t^*)=\max_{n\in S}\{\mathcal{R}[W_1]\}-\min_{n\in S}\{\mathcal{R}[W_1]\}.$ Note that \citet{lewis2} obtained $f(t^*)=2$ since their space variable was continuous. However when the space variable is discrete, it is not always possible to obtain this equality. In the following lines, we find the lower and upper bounds for $f(t^*)$ for the parameters given in examples \textbf{E1-2}.

\begin{figure}[h!]

\centering
\subfigure[]{
\includegraphics[width=7cm, height=7cm]{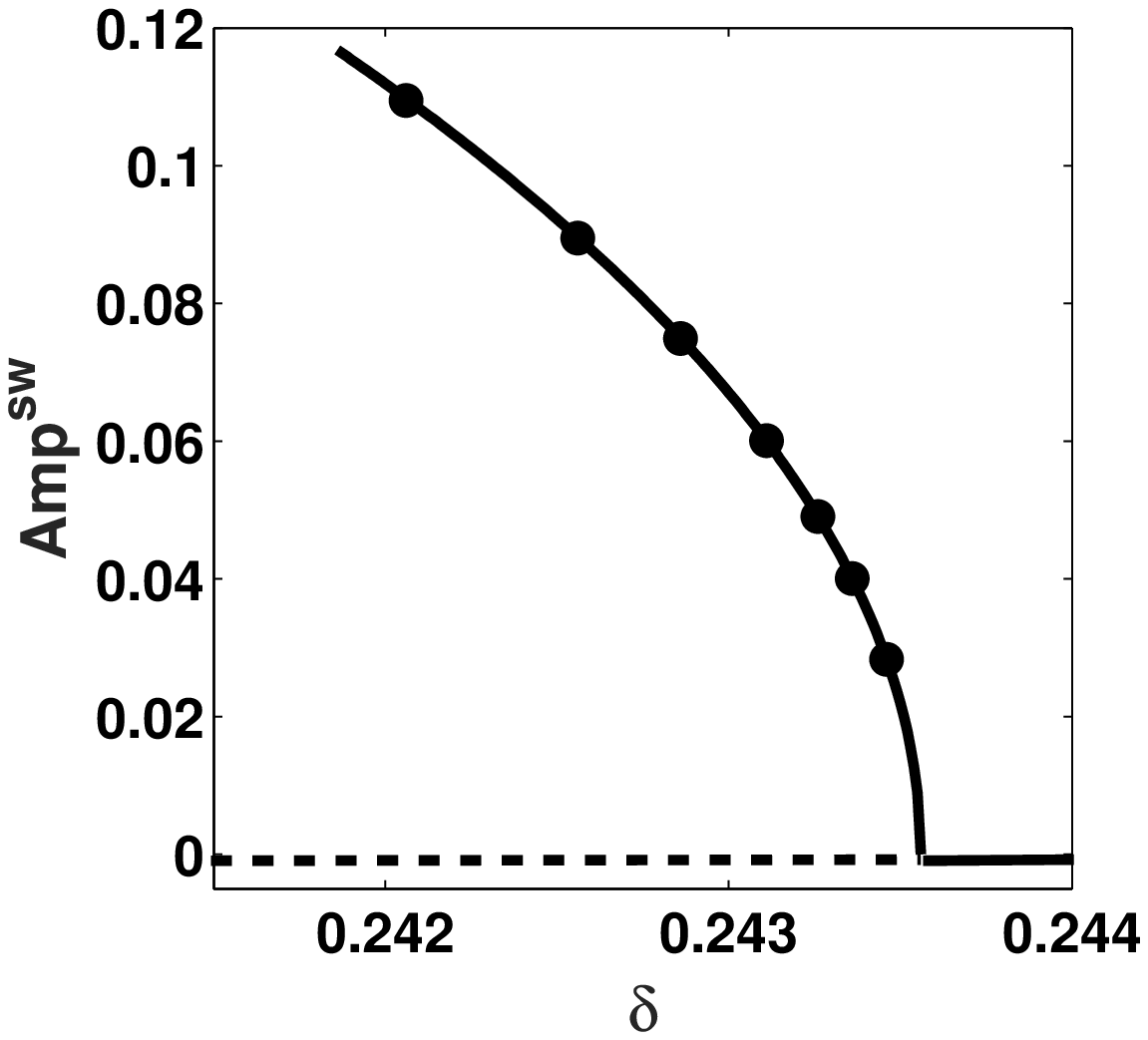}
}
\subfigure[]{
\includegraphics[width=7cm, height=7cm]{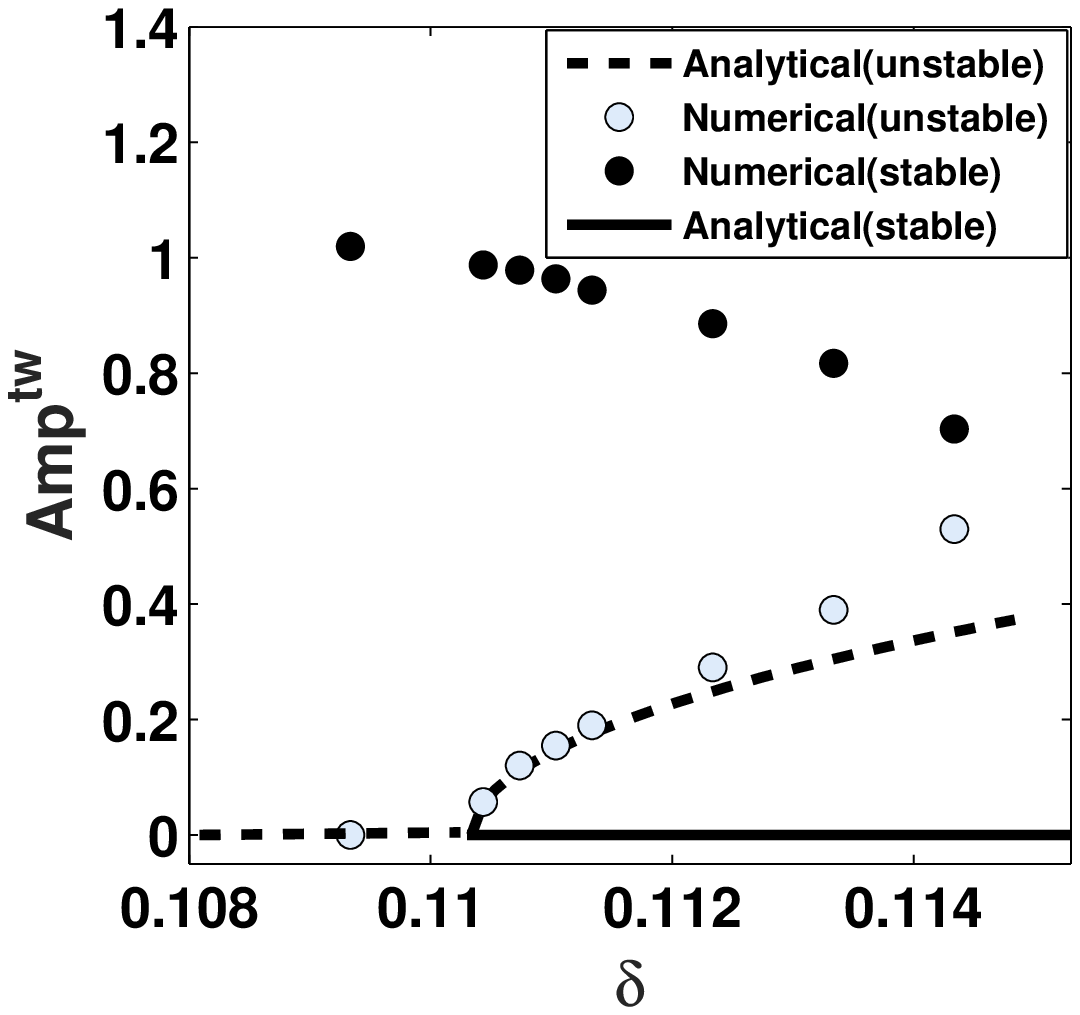}
} 
\caption{Amplitude levels of stationary and traveling waves near critical dispersal rates. Panel (a) illustrates amplitude levels of stationary waves for the kernels and initial condition described in \textbf{E1}. Similarly, panel (b) shows the amplitude levels of traveling waves for the kernels and initial condition described in \textbf{E2}. }
\label{ampl}
\end{figure}

Now consider the kernels described in examples \textbf{E1-2} again. Figure \ref{ampl} (a) shows the stable branch (black curve) obtained by using the analytical results and stable pattern amplitudes (black dots) for small perturbations of critical nonlocal dispersal rate $\delta_0.$ {As expected, we observe that decreasing the dispersal rate increases the amplitude of the patterns (or cluster sizes).} As seen from the figure, nonlinear prediction \eqref{amplitude} and the amplitude obtained from numerical simulations of \eqref{int} stays pretty close for small values of $\veps.$ Here we take $f(t^*)=1.9830.$ Note also that this value does not depend on $t^*,$ since the eigenvalues are real {\it i.e.} $w_0=0.$

Figure \ref{ampl} (b) reveals that the subcritical branch of periodic solutions connects to
the spatially inhomogeneous solutions at the touchdown solution. The subcritical branch of amplitude levels of spatial patterns turns around, and eventually terminates at the spatially homogenous solution $\mathbf p=\mathbf 1.$ From the bifurcation diagram, we see that as the dispersal rate $\delta$ is decreased the spatially homogenous state destabilizes, the transition will involve a
jump to a large-amplitude pattern. { One interesting biological prediction which follows
is that of hysteretic transitions for the amplitudes of solutions. Populations with high enough mobility rates
will not form patterns. If the dispersal rate is slightly larger than the critical one, the unstable branch determines a threshold value for the amplitudes of the initial densities. We observe patterns amplitude levels of which are obtained numerically (black dots) if the amplitude of the initial densities passes this threshold value. On the other hand, the spatial inhomogeneity disappears when the amplitude of the initial density is below this threshold. For smaller dispersal rates this threshold vanishes and spatial inhomogeneity is observed for any spatial perturbation of homogenous state.} 

We would like to note that the cubic S-L equation is enough to characterize possible bifurcations when the system parameters are in hand. In particular, the parameter $\mathcal R[\Psi]$ is the key to finding the type of the bifurcation as described in Section \ref{analytical}. As an example, consider the following parameter set: $\mathbf d=\mathbf u^{(2,0)}$ and $\mathbf c=\mathbf{u}^{(r,a)}$ for $r\in \{1,2,\cdots,8\}$ and $ a\in \{0,1,\cdots8\}.$ We find that the subcritical bifurcations can be observed only for values $(r,a)$ in $\{(1,5),(3,4),(4,2),(4,4),(5,2),(7,3),(8,7),(8,8)\}.$ 

\subsection{Approximating amplitudes for subcritical instabilities}\label{subcr}

{To further analyze} the model, we need to consider higher order correction terms. {To accomplish} such a task, we take the complex amplitude as a function of two time scales $ A(\tau,\tau_1)$ where the second time scale is given by $\tau_1=\veps^4t$ and take the bifurcation parameter $\delta$ as 

\beq \delta=\delta_0+\mu(\veps^2+\veps^4).\feq 

In the case of discontinuous transitions to instability, we need to expand the equations up to fifth order. The calculations are given explicitly in Appendix \ref{ap}. A quintic equation for the partial derivative $A_{\tau_1}$ is given by \eqref{qsl}. Using this equation along with the equation \eqref{sl}, one can obtain the following quintic S-L equation:

 \beqn\label{tsl} \frac{dA}{d\tau}=A_\tau+\veps^2A_{\tau_1}=\mu\Phi(1+\veps^2)A+(\Psi+\veps^2\Theta_3)A|A|^2+\veps^2\Theta_5|A|^4A.
\feqn 

For the derivation of this equation see Appendix \ref{ap} where we also give explicit expressions for parameter values $\Theta_3$ and $\Theta_5$ (see formulas \eqref{t3} and \eqref{t5}). Note also that this equation is only valid if both of the Assumptions \ref{a1} and \ref{a2}, given in Appendices \ref{cubic} and \ref{ap}, hold.
Again the amplitude function $A$ is complex valued. Hence we write it as $A(\tau)=L(\tau)e^{i\theta(\tau)}$ with real terms $L(\tau)=|A(\tau)|$ and $\theta(\tau).$ Plugging this equality in quintic S-L equation \eqref{tsl} one obtains:
\begin{flalign}
\label{tli}
\frac{dL}{d\tau}&=L\Bigr(\mu(1+\veps^2)\mathcal{R}[\Phi]+\mathcal{R}[\Psi+\veps^2\Theta_3]L^2+\veps^2 \mathcal{R}[\Theta_5]L^4\Bigl),\\
\frac{d\theta}{d\tau}&=\mu(1+\veps^2)\mathcal{I}[\Phi]+\mathcal{I}[\Psi+\veps^2\Theta_3]L^2+\veps^2 \mathcal{I}[\Theta_5]L^4. \nonumber
\end{flalign}

We are interested in the real roots of RHS of equation \eqref{tli}. Equilibrium $L_0=0$ is again unstable for $\mu=-1$ and it is stable for $\mu=1.$ 

The RHS of \eqref{tli} has four roots other than zero determined by 

\beqn\label{root}{ L^2}=-\frac{\mathcal{R}[\Psi+\veps^2\Theta_3]\pm\Bigl(\mathcal{R}[\Psi+\veps^2\Theta_3]^2-4\veps^2\mu(1+\veps^2)\mathcal{R}[\Phi]\mathcal{R}[\Theta_5]\Bigr)^{1/2}}{2\veps^2 \mathcal{R}[\Theta_5]}\feqn

Hence we have the following classifications regarding the equilibria of \eqref{tli} and approximations to the amplitudes of the patterns.

{\it Case 1:} Suppose that $\mu=1.$ Then \eqref{tli} has four real roots other than zero and these roots are symmetric around zero. Consider only the positive roots and denote them by $L_+$ and $L_-.$ Then the unstable branch can be approximated by $\veps L_-$ to the right of $\delta_0.$ In addition, $\veps L_+$ approximates the stable branch.

{\it Case 2:} Suppose that $\mu=-1.$ Then \eqref{tli} has only two real roots other than zero which are symmetric around zero. There is no unstable branch here, and one can approximate the stable branch by $\veps L_+.$

\begin{remark}\label{rem2}
It is clear from the formula \eqref{root} that $L_+\in O(1/\veps).$ This implies that the stable branch approximated by $\veps L_+$ that is of order $1$ and the magnitude of error term for the approximation is of order $\veps.$ 
\end{remark}

\begin{figure}[h!]

\centering
\subfigure[]{
\includegraphics[width=7cm, height=7cm]{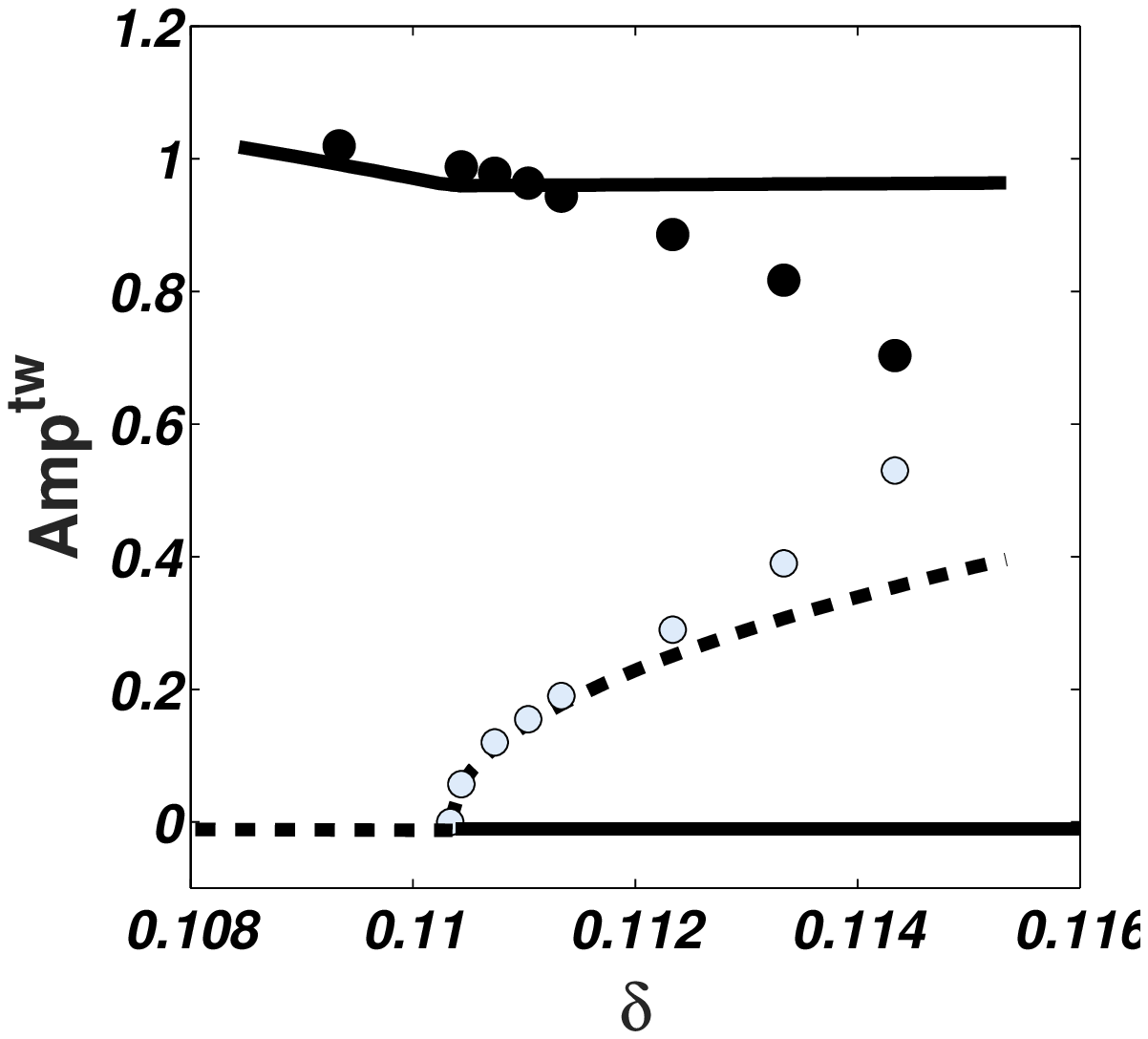}
}
\subfigure[]{
\includegraphics[width=7cm, height=7.1cm]{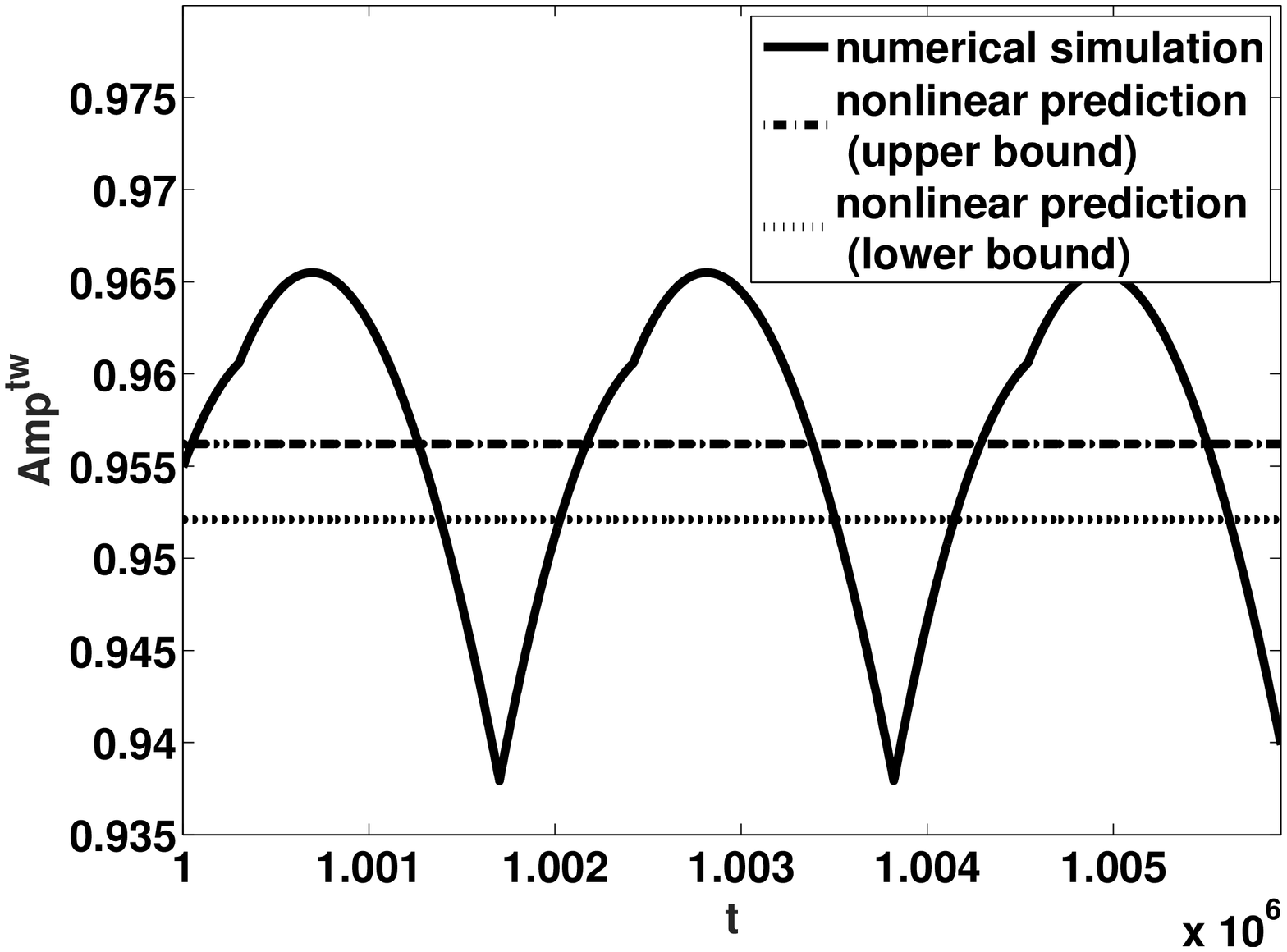}
}
\caption{Panel (a) shows the approximations to stable and unstable branches of amplitude levels of patterns with parameters given in \textbf{E2}. In Panel (b), a plot of amplitude levels of the traveling waves for the dispersal rate $\delta=0.1104$ is illustrated. The dashed curves shows the upper and lower bounds of amplitude levels predicted by weakly nonlinear analysis.}
\label{qq}
\end{figure}

{The unstable branch (dashed curve) shown in Figure \ref{qq} Panel (a) is calculated using the root $L_-=2.8046$ rather than $L_1=2.7155$ for the parameters given in \textbf{E2}.} The curve again determines the maximum amplitude of initial conditions below which the space homogenous equilibrium $L_ 0$ is reached. If the amplitude of the initial condition passes this maximum value then one can observe pattern formation whose amplitude levels are obtained numerically (black dots). An approximation of stable amplitude levels is given by the black curve for $f(t^*)=1.99$. Note that the approximation stays relatively close to numerically obtained amplitude levels (black dots) when $\delta$ is close enough to $\delta_0.$ Yet as the distance between $\delta$ and $\delta_0$ increases, the approximation fails  to capture the numerically obtained amplitude levels since the magnitude of error term is given by  $O\Bigl(\sqrt{|\delta-\delta_0|}\Bigr)$ (see Remark \ref{rem2}). {Quintic SL equation mainly predicts the maximum difference between the densities of the residents living in different patches. Hence this approximation informs us about the degree of spatial inhomogeneity. Another observation is that decreasing the dispersal rate leads more denser group of individuals as in the case of supercritical bifurcations. }

 In Panel (b), the actual amplitude levels of the patterns are plotted as a function of time $t$ for dispersal rate $\delta=0.1104.$ The upper and lower bounds are calculated using weakly nonlinear predictions for the values $f(t^*)=1.9830 \text{ and }1.9915,$ respectively. 

\section{Discussion} \label{conc}

There are many models for spatial pattern formation in biology literature (for reviews, see \cite{murray,okubo}). Many of these models consist of coupled
systems of PDEs and other continuum models. {For these models, it was shown that diffusion-driven instabilities are possible and can result in the exhibition of a vast range of spatial patterns.} In this paper, we have considered a single-species population model with nonlocal competition in a discontinuous habitat. This model is a  continuous in time and discrete in space analog of the nonlocal Fisher equation \citep{pattern} and also the spatial Beverton-Holt model \citep{doebelli}. We considered our spatial habitat as a finite lattice with wrap around boundary conditions. A few simplifying but biologically reasonable assumptions are made to model the nonlocal interactions and dispersal (see Section \ref{modelnotation}). As a result, we have developed a model for ecological pattern formation consisting of a system of coupled  ordinary differential equations.

 Taking into account research conducted by \cite{main}, nonlocal competition kernel models the consumption of resources in a patch by an individual breeding in some other site. {In addition, passive dispersal term models the dispersal of conspecifics from their natal or breeding sites.} To show the system has spatially heterogeneous solutions, we performed a linear stability analysis. Provided that the dispersal rate is sufficiently small, such heterogeneous solutions exist for a class of competition kernels satisfying certain conditions.  We also numerically  studied these patterns near critical dispersal rate and illustrated numerical simulations of traveling and stationary wave type patterns in Figure \ref{stationary}.

  Note that small dispersal rate and the condition on the competition kernel (real part of DFS of it takes negative values) are needed to observe pattern formation. These two requirements together imply that dispersal itself does not lead to pattern formation in single-species models; and hence, the nonlocal competition term is crucial to observe spatial or spatio-temporal patterns. In addition one can conclude that the high mobility of individuals, {\it i.e.}  a large dispersal rate, stabilizes the space homogenous solution $\mathbf1$ and spatial inhomogeneity vanishes. Such a phenomenon is extensively discussed in the literature (see, e.g. \cite{hastings1,doebeli2,mig}). 

Despite weakly nonlinear analysis is  being widely applied to continuum models \citep{murray,lewis2,lewis1}, to the best of our knowledge weakly nonlinear analysis of a metapopulation model has not been studied previously. As a result of this analysis, we concluded that our cubic S-L equation characterizes the type of the bifurcations depending on the system parameters. In the case of supercritical bifurcations, the stable equilibrium of S-L equations can be used to approximate amplitudes of the patterns. However, for subcritical bifurcations, one can only obtain the unstable branch using the cubic S-L equation. 

{Spatial models are often used to address general questions  concerning the effects of spatial structure on the nature and complexity of population fluctuations \citep{doebelli}. Amplitudes as defined by \eqref{amp} can be considered as a measure of these fluctuations. Hence, it is important to approximate the amplitude levels of patterns.} As suggested by \citet{lewis2}, we added higher-order terms to obtain a quintic S-L equation allowing us to approximate the amplitudes of the patterns in the case of subcritical bifurcations.  Hence, amplitude levels of patterns are approximated using cubic and quintic S-L equations for supercritical and subcritical bifurcations, respectively.

In the case of both subcritical and supercritical bifurcations, we  observed that the spatial inhomogeneity becomes more apparent when the mobility of individuals is reduced. This is to say that the amplitudes of patterns increases as the dispersal rate decreases. The cubic equation suggests that there is a threshold population density above which one can observe pattern formation. In addition below this threshold spatial inhomogeneity is not apparent anymore.

Our results can be summarized as follows:

\begin{itemize}
\item[(1)] We presented a metapopulation model taking nonlocal interactions and dispersal into account that exhibits stationary and traveling wave type patterns near the stability boundaries. { When asymmetric kernels are used to model nonlocal interactions or diffusion, the resulting patterns are of travelling wave type referring to a non-uniform distribution on habitat patches, in which the population density varies in one spatial direction, as well as in time.  These spatial and temporal oscillations together give rise to the appearance of a wave in population density (see Figure \ref{stationary}(b)). The stationary wave type patterns arise as a result of symmetric nonlocal interactions and dispersal. Such patterns can also be identified as a non-uniform, time-invariant distribution of population densities on habitat patches (see Figure \ref{stationary}(a)).}
\item[(2)] We  used weakly nonlinear analysis to study phase transitions between different behaviors. We found that the both continuous and discontinuous transitions between ordered and disordered behaviors are possible. {The subcritical bifurcation suggests that there is a threshold amplitude for the population density, such that amplitudes below this threshold will disperse, while amplitudes
above this threshold will become even more apparent and persist for a longer time. This bifurcation describes the 
transition between the spatial homogeneity,
and the cluster formation represented by the high-density stationary or traveling
wave type patterns. Whereas supercritical bifurcation suggests that there is a continious transition between spatial homogeneity and the cluster formation whose amplitudes are relatively small near the stability boundaries.}
\item[(3)] The amplitudes of the patterns for supercritical and subcritical bifurcations were approximated using the stable nonzero equilibrium of cubic and quintic S-L equations. {While cubic S-L equation completely characterizes the supercritical instabilities near critical dispersal rate it cannot predict the amplitudes of the patterns when subcritical bifurcations arises. We obtained the quintic S-L equation to get an approximation for the maximum difference in densities of groups when transition to instability is discontinuous. We also observed that smaller dispersal rate results in more denser groups of individuals for both sub- and super-critical bifurcations.}
\end{itemize}

One can take foraging organization in Pharaoh's ants as a biological example exhibiting continuous and discontinuous transitions between disordered and ordered behaviors. \citet{ant} show that small colonies of these ants forage in a disorganized
manner, yet this behavior evolves to organized pheromone-based foraging
in larger colonies. They also show that the transition between disorganized and organized foraging exhibits hysteresis ({\it i.e.} subcritical bifurcations) when food sources are difficult to locate. {Their study shows that colonies of 600 ants or less are unable to forage in an organized manner to a faraway feeder. Hence the findings of \citet{ant} suggests that the nonlocal competition distance for Pharaoh's ants living in a site depends on the number of individuals residing in that patch. To obtain a more realistic model, one can extend the model proposed and investigated in this study by taking a density dependent nonlocal competition kernel into account. }

Finally there are some interesting issues that should be further explored and extended. In this paper, we focused on the equation \eqref{int} in a one dimensional habitat. Weakly nonlinear analysis can be extended to two or three dimensional lattices. Weakly nonlinear analysis as presented in this paper is not used to study any other metapopulation model. Hence, we believe that it can be used to understand the behavior of many other metapopulation models. The most apparent example might be in employing this method for a metapopulation model with dispersal driven instabilities (see {\it e.g.} \citet{ecocom}).

\begin{acknowledgement}
The author thanks the anonymous reviewer for his/her careful reading of this manuscript and his/her many
insightful comments and suggestions that improved the manuscript.
\end{acknowledgement}
\appendix

\section{DFS and Linear Analysis}
\subsection{Calculation of DFS for a class of uniform kernels}\label{dfsu}

 Using the periodicity of $\mathbf u^{(r,a)}$ and complex exponentials, one can calculate $ U_k$ for any nonzero integer $k\in S$ as follows:\beq  U_k &=&\sum_{n=0}^{r+a}u_n^{(r,a)}e^{-2j\pi nk/N}+ \sum_{n=r+a-N+1}^{-1}u_{(n)_N}^{(r,a)}e^{-2j\pi nk/N}\\
&=&\frac{1}{2r+1}\sum_{n=-r+a}^{r+a}e^{-2j\pi nk/N}\\
&=&\frac{e^{-2j\pi n(a-r)/N}}{2r+1}\sum_{l=0}^{2r}e^{-2j\pi nl/N}\\
&=&\frac{1}{2r+1}\frac{e^{-2j\pi k(a-r)/N}-e^{-2j\pi k(a+r+1)/N}}{1-e^{-2j\pi k/N}}.
 \feq Clearly $ U_0=1$ from the first equality. We would like yo note that the characteristic function of discrete uniform distribution with support  $\{a-r,a-r+1,\cdots, r+a\}$ is given by \beq \mathcal C(t)=\frac{e^{j(a-r)t}-e^{j(a+r+1)t}}{(2r+1)(1-e^{jt})}.\feq Hence above given formula can be obtained from the characteristic  equation by evaluating it at discrete values $t=-2\pi k/N$ for $k\in S.$ It is also possible to use other discrete distributions and compute their DFS by using the shift theorem and their characteristic functions \citep{dft,mandal}. 
 
 \subsection{Discussion regarding Assumption \ref{ass}}\label{discussion}
 
 To show Assumption \ref{ass} is not redundant, it is enough to construct an example for which the state $\mathbf 1$ is unstable for any $\delta>0.$ Consider a habitat with 8 patches {\it i.e.} $N=8.$ Take the dispersal kernel $\mathbf d$ with $d_2=d_{<-2>_8}=d_6=\frac{1}{2}.$ The DFS of this vector is given by $\mathbf D=( 1  ,0 ,-1 ,0 ,1  ,0 ,-1 ,0).$ Note that $D_4=1.$ Consider also uniform kernel $\mathbf u^{(1,0)}$ as the interaction kernel. One can calculate the DFS of this kernel as $\mathbf C=(1.0000 ,   0.8047 ,   0.3333 ,  -0.1381 ,  -0.3333   -0.1381 ,   0.3333 ,   0.8047).$  This implies one of the eigenvalues of the coupled ODE system $\lambda(\delta,4)=0.3333$ and hence the state is unstable for any $\delta>0.$

\section{Derivation of Cubic S-L Equation}\label{cubic}

Before proceeding the perturbation analysis, define the following complex constants \beq \mathcal L_{\delta_0}^n= njw_0-\delta_0(D_{<nk_c>_N}-1)+C_{<nk_c>}\feq for $n \in S.$ In addition, note that we need the following technical assumption to be able to obtain cubic S-L equation. 

\begin{assume}\label{a1}
We assume that $k_c\neq0$ and $<4k_c>_N\neq0$ regarding the system parameters $k_c$ and $N.$
\end{assume}

Plugging the solution \eqref{psoln} into \eqref{int} gives the following relation at level $O(\veps):$ \beq A(\tau)\mathcal L_{\delta_0}^1W_1=0.\feq Hence at this level one obtains the linear relation \eqref{eigen} and it is not possible to gather information about the amplitude $A.$

At level $O(\veps^2)$ we obtain the following equation: \beq \mathcal L_{\delta_0}^0A_{20}W_0+( \mathcal L_{\delta_0}^2A_{22}W_2+c.c.)=\gamma_{20}AA^cW_0+\bigr( \gamma_{22}A^2W_2 +c.c\bigl)\feq where $ \gamma_{20}=-2\mathcal R[C_{<k_c>_N}]$ and $\gamma_{22}=-C_{<k_c>_N}.$  

 By Assumption \ref{a1}, it is easy to see that $W_0,$ and $W_{\pm2}$ are linearly independent vectors ( $<4k_c>_N\neq0$ implies $<2k_c>_N\neq0$). Hence we have : \beq A_{20}&=&\gamma_{20}AA^c\\
 A_{22}&=&\frac{\gamma_{22}}{\mathcal L_{\delta_0}^2}A^2.\feq Note also that Assumption \ref{a1} guarantees that $W_2$ is not in the kernel of the linearized operator so that $\mathcal L_{\delta_0}^2\neq0.$
 
 At the level $O(\veps^3)$ we obtain the following equality\beqn\label{solv}\mathcal L_{\delta_0}^3A_{33}W_3+c.c.=\bigl(-A_\tau(\tau)+\mu\Phi A+\Psi A^2A^c \bigr)W_1+ \gamma_{33} A^3+c.c.\feqn where \beqn\label{phi}\Phi&=& (D_{<k_c>_N}-1),\\ \label{psi}~~\Psi&=&-\gamma_{20}(1+C_{<k_c>_N})-\frac{\gamma_{22}}{\mathcal L_{\delta_0}^2}(C_{<k_c>_N}^c+C_{<2k_c>_N})\\ \text{ and  }\nonumber \\ \label{g33}\gamma_{33}&=&-\frac{\gamma_{22}}{\mathcal L_{\delta_0}^2}(C_{<k_c>_N}+C_{<2k_c>_N}).\feqn By Assumption \ref{a1}, one can easily see that $<3k_c>_N\neq\pm k_c.$  Hence, we can easily conclude that the coefficient of $W_1$ is zero in \eqref{solv}, from which we get quintic S-L equation \eqref{sl}.

\section{Derivation of Quintic S-L Equation}

\label{ap}

To obtain quintic S-L equation, we need more restrictions. Here we suppose that Assumption \ref{a1} holds in addition to the following restrictions.
\begin{assume}\label{a2}
We assume that $<6k_c>_N\neq0$ regarding the system parameters $k_c$ and $N.$ 
\end{assume}

Taking in to account that the cubic S-L equation \eqref{sl} is still valid for complex amplitude $A$, we take the solution at level $O(\veps^3)$ as follows \beqn \label{3rd} \mathbf p^{(3)}=\gamma_{3} A^3 W_3 +c.c\feqn where $\gamma_3=\frac{\gamma_{33}}{\mathcal L_{\delta_0}^3}.$ In addition, for the sake of simplicity, we denote $\frac{\gamma_{22}}{\mathcal L_{\delta_0}^2}$ by $\gamma_2.$

 Here we consider an expansion of solution $\mathbf p$ as follows:  \beqn\label{ansatz}
\mathbf p=1+\sum_{m=1}^{4}\veps^m \mathbf p^{(m)} (t,\tau,\tau_1)+O(\veps^5) \feqn where $\mathbf p^{(i)}$ for $i=1,2,3$ are as defined in Section \ref{analytical} and the fourth term has the following form:  \beq \mathbf p^{(4)}=A_{40}W_0+(A_{42}W_2+A_{44}W_4 +c.c.).\feq Plugging this solution into equation \eqref{int} one obtains $A_{20}$ and $A_{22}$ as given in Appendix \ref{a1}. The solution $\mathbf p^{(3)}$ is given by \eqref{3rd}. Hence we need to determine $\mathbf p^{(4)}.$   At levels $O(\veps^4)W_i$ for $i=,0,2,4$ we get the following equalities:
 \beq A_{40} =\gamma_{40}^4 |A|^4 +\gamma_{40}^2 |A|^2 \text{ and } A_{42} =\gamma_{42}^3A| A|^2 +\gamma_{42}^2 A^2\feq 
 where \beq\gamma_{40}^4&=&-(\gamma_{20}^2 + \gamma_{20}(\Psi^c + \Psi) + \gamma_2\gamma_2^c(C_{<2k_c>_N} + C_{<2k_c>_N}^c))\\ \gamma_{40}^2&=&- \mu\gamma_{20} (\Phi +  \Phi^c)\\ \gamma_{42}^3&=&  -\big(\gamma_3(C_{<3k_c>_N} + C_{<k_c>_N}^c) + \gamma_2(\gamma_{20}(1+ C_{<2k_c>_N})+ 2 \Psi)\big)/\mathcal L_{\delta_0}^2\\ \gamma_{42}^2 &=&-\gamma_2 (\mu(1-D_{<2k_c>_N}) + 2\Phi)/\mathcal L_{\delta_0}^2.\feq
 Note that function $A_{44}$ does not contribute at level $O(\veps^5)W_1.$ Above equalities requires the linear independency of vectors $W_0,$ $W_{\pm2}$ and $W_\pm4$ which follows from Assumptions \ref{a1} and \ref{a2}. Hence, at the level $O(\veps^5)W_1$, we have the following equation:\beqn\label{qsl}A_{\tau_1} - \mu\Phi A  + \Theta_3A| A|^2 + \Theta_5A| A|^4=0
\feqn
 where  \beqn\label{t3} \Theta_3=\gamma_{40}^2(1+C_{<k_c>_N}) + (\gamma_{42}^2+\gamma_{42}^3)(C_{<k_c>_N}^c+C_{<2k_c>_N}) \feqn and \beqn\label{t5}\Theta_5=\gamma_{40}^4(C_{<k_c>_N} + 1) +\gamma_{3}\gamma_2^c(C_{<3k_c>_N} + C_{<2k_c>_N}^c).\feqn Note that this equation is also obtained from the fact that the vectors $W_\pm1$ and  $W_5$ are linearly independent.
 
 Hence, by combining equations \eqref{sl} and \eqref{qsl}, one obtains quintic S-L equation \eqref{tsl}
                  

%
%

\end{document}